# Designing Pricing Schemes based on Progressive Tariff and Consumer Grouping in Migration to a Future Smart Grid

Kyeong Soo Kim
Department of Electrical and Electronic Engineering, Xi'an Jiaotong-Liverpool University, Suzhou, China
Center for Smart Grid and Information Convergence (CeSGIC), XJTLU, Suzhou, China
Kyeongsoo.Kim@xjtlu.edu.cn

*Abstract*—We study the design of pricing schemes for a group of consumers with smart meters (e.g., in a Greenfield area) who are connected through a gateway to a traditional electricity greed with a progressive tariff. Because the progressive tariff cannot take into account the time aspect of electricity demands, we apply it to consumers in both an individual and a group basis over a shorter time period, which can flatten the overall demand over time and thereby reduce peak load. This scenario for the coexistence of traditional and smart girds and the pricing schemes under this scenario can enable smooth migration to a future smart grid.

## I. INTRODUCTION

Smooth migration from a current electricity grid to a future smart grid has been the major focus of researches in Academia, Industry, and Government: The large investments needed for the augmentation of power grid and the installation of two-way communication infrastructures could be a major barrier for roll-out of smart grid by power utilities, while the privacy issues related with the power-usage data exchanged between the consumer and the power utility has become a major concern for consumers in adopting smart grid technologies [1]. To overcome these barriers and issues in migrating to a smart grid, the UK government outlines three phases of smart grid development [2]. Arian et al. separate smart metering and smart grid infrastructures and suggest intelligent migration strategies from smart metering to smart grid [3].

These works clearly show that the migration to a future smart grid should take a step-wise or phased approach and that it may take long time to completely replace a traditional grid with a smart grid in a national scale. This means that we need to prepare for the coexistence of a traditional grid and a smart grid for a long period of time in the migration to a future smart grid.

In this paper we take one of coexistence scenarios, where a group of consumers with smart meters (e.g., in a Greenfield area) are connected through a gateway to a traditional electricity greed with a progressive tariff, and study how to design pricing schemes based on the progressive tariff that can flatten the overall demand over time and thereby reduce peak load.

## II. DESIGNING PRICING SCHEMES BASED ON PROGRESSIVE TARIFF AND CONSUMER GROUPING

### A. A System Model

Fig. 1 shows a system model for the coexistence of a traditional electricity grid and a smart grid, which we base our discussions on. The gateway, connecting the two grids, receives electricity and progressive tariff information from the traditional grid, while interacting with smart meters at customer premises through the bi-directional communication links for efficient power distribution and intelligent billing based on real-time usage monitoring. From the point of view of the traditional grid, the gateway functions as a traditional electricity meter for a virtual consumer who represents the group of consumers with smart meters; from the point of view of the smart grid, on the other hand, the gateway works as a utility control center in a typical smart grid [4].

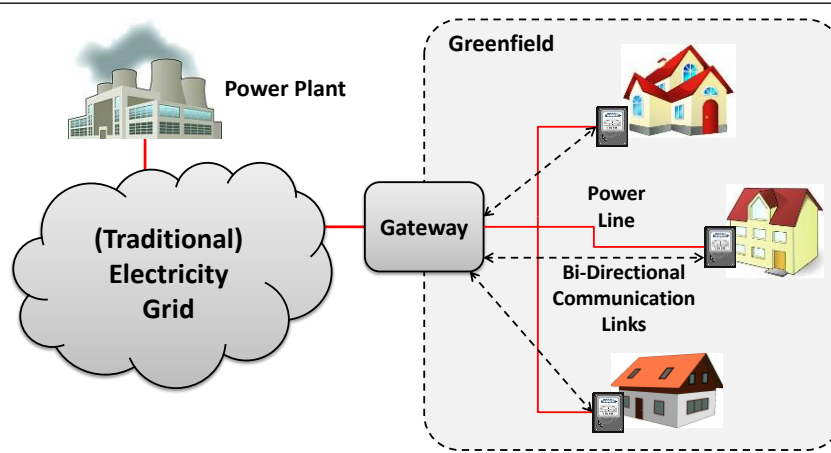

Figure 1. A system model for the coexistance of a traditional electricity grid and a smart grid.

Under a progressive tariff, a consumer's electricity bill is calculated based on multiple progressive levels according to its power consumption [5]. For instance, consider the monthly progressive tariff for residential consumers by Korea Electric Power Corporation (KEPCO) summarized in Table I [6].

TABLE I. ELECTRIC RATES FOR RESIDENTIAL (LOW-VOLTAGE) CUSTOMERS BY KEPCO [6]

| Consumption Range [kWh] | | Energy Charge (KRW[a]/kWh) |
|---|---|---|
| Tier 1 | ~ 100 | 60.7 |
| Tier 2 | 201-200 | 125.9 |
| Tier 3 | 201-300 | 187.9 |
| Tier 4 | 301-400 | 280.6 |
| Tier 5 | 401-500 | 417.7 |
| Tier 6 | 500 ~ | 709.5 |

a. Korean Won

If monthly power consumption for a family is 350 kWh, the price (in KRW) is determined as follows:[1]

$$100 \cdot 60.7 + 100 \cdot 125.9 + 100 \cdot 187.9 + 50 \cdot 280.6 = 51480$$

Unlike advanced pricing schemes based on time-based metering (e.g., time-of-use (TOU) price scheme [5]), the progressive tariff cannot take into account the dynamic nature of usage patterns in a shorter time scale (e.g., an hour). The progressive tariff, however, can provide incentives for consumers to change their usage patterns over the whole billing period (e.g., a month) by rewarding savings and penalizing higher consumption [7], which is the reason it becomes popular as a residential pricing scheme in several countries including Japan, Korea, Taiwan, and the US [8].

### B. Progressive Tariff over Time Slots with Consumer Groupting

Smart grid technologies provide advanced real-time usage monitoring based on smart meters and the interaction between the utility and consumers through bi-directional communication links in determining actual consumption of electricity. Based on such capabilities, the utility can offer dynamic and cost-reflective tariffs which take into account not only the consumption of electricity by consumers but also the generation of electricity by the utility and/or distributed energy sources in a microgrid [2], [4]. In the coexistence scenario described in Sec. II.A, however, we cannot expect such dynamic tariff information from the utility in the traditional grid; instead, the gateway should provide an internal pricing scheme for the consumers in the smart grid based on the external, static pricing scheme (i.e., progressive tariff) from the utility.

Here we discuss how to enable the design of such a dynamic pricing scheme for the internal smart grid. Fig. 2 illustrates two key ideas — i.e. (1) the adaptation of the progressive tariff for a shorter time period (i.e., 6-hour time slots in these examples) and (2) grouping of consumers in applying the progressive tariff — through daily usage pattern examples of two consumers and the monthly progressive tariff by KEPCO described in Table I.

Note that the energy consumption range of each tier has been adjusted according to the change in a time period. For instance, the maximum value for Tier 1 needs to be scaled down for a 6-hour time slot as follows (assuming 30 days per month):

$$100 \times \frac{1}{30} \times \frac{6}{24} = 0.8333 \ (kWh)$$

Fig. 2 (a) and (b) show an original and a modified daily usage patterns of two consumers, respectively. If we apply the progressive tariff adjusted for a 6-hour time slot to each consumer individually (i.e., the graphs on the left side of the figure), there is no incentive for demand response (i.e., shifting the usage to other time slots) from Consumer 2 because the structure of the progressive tariff (i.e., the range and the energy charge for each tier) is static and not varying over time depending on the consumption and the supply of electricity. In fact, assuming that the same daily usage patterns repeat over the whole month, the prices for the usage patterns shown in Fig. 2 under the progressive tariff over the 6-hour time slot are higher than those under the original monthly progressive tariff: For the former, the rates for Tier 1 through 3 are used in price calculation, while for the latter, only the rate for Tier 1 is used.

Once we group the consumers together and apply the progressive tariff to the group in a collective manner[2], however, we can provide incentives for demand response. Fig. 2 (b) shows that usages of Consumer 2 for two consecutive time slots belong to Tier 1 by consumer grouping, while they spread over both Tier 1 and 2 without such grouping. This benefit of consumer grouping is similar to that of resource sharing in networking area [9]: The unused resource by inactive users (i.e., the amount of electricity belonging to Tier 1 which is not used by Consumer 1) is reallocated to active users (i.e., Consumer 2) through resource sharing.

### C. Allocation of Group Price to Individual Consumers

Under the consumer grouping, determining the prices for individual consumers becomes an important issue: Because the group price for a given time slot under the consumer grouping is not equal to the sum of individual prices in general, it is a challenge how to allocate this group price to each consumer. Fig. 3 illustrates the issue of price allocation under the consumer grouping, which compares the total prices of individual and group pricing schemes for three consumers for a 6-hour time slot. As shown in the figure, the total price under the group pricing scheme (i.e., 466.50 KRW) is less than that of the individual pricing scheme (i.e., 518.16 KRW).

---

[1] Standing and demand charges are excluded for simplicity.

[2] Compared to the ranges for the individual application, the ranges are scaled up this time by the number of consumers (i.e., 2).

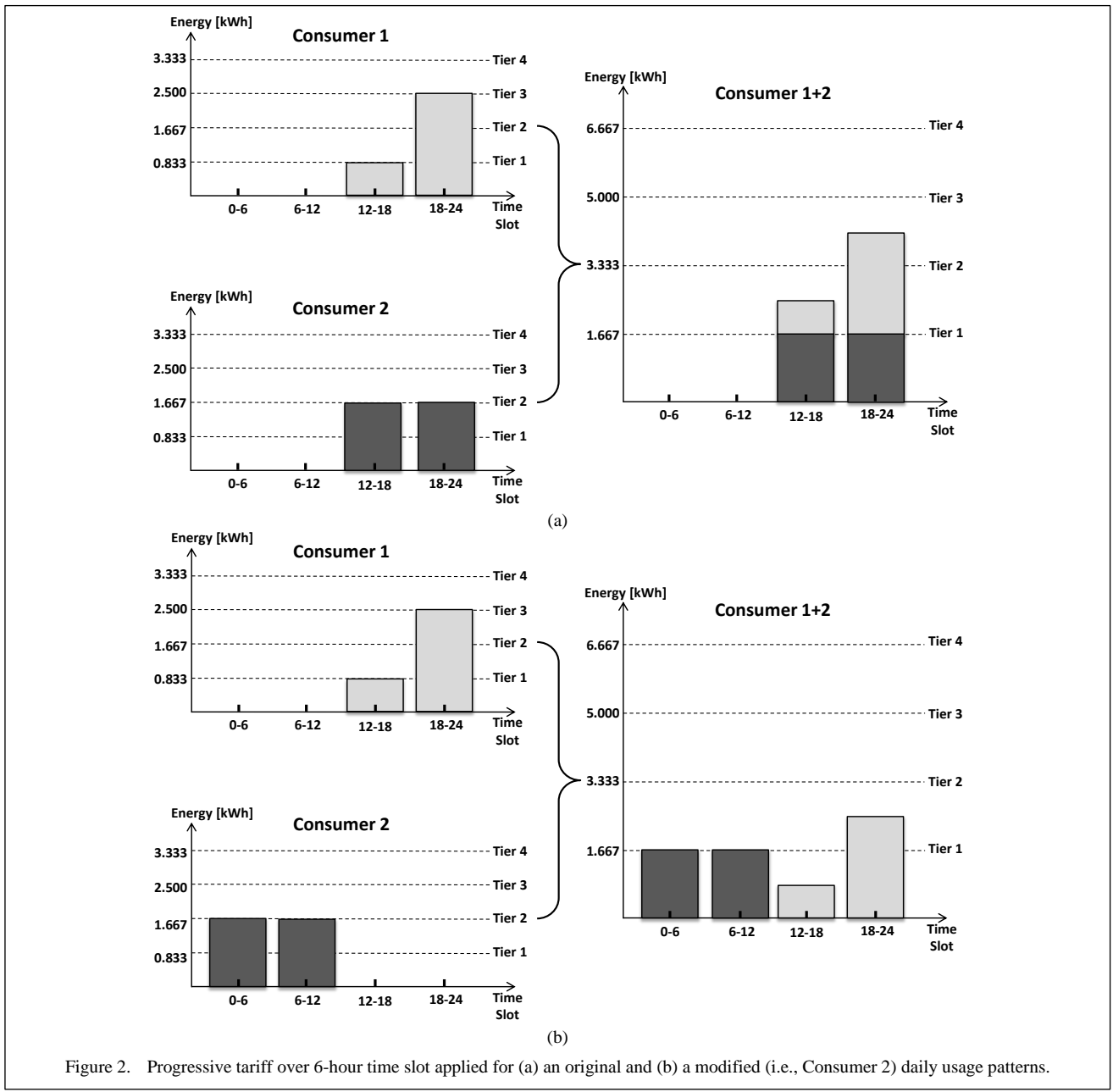

Figure 2. Progressive tariff over 6-hour time slot applied for (a) an original and (b) a modified (i.e., Consumer 2) daily usage patterns.

One possible allocation strategy is a proportional allocation (e.g., as in [10]), where the total price under the group pricing scheme is allocated to consumers proportional to their individual prices under the individual pricing scheme. According to the proportional allocation, the price (in KRW) for each consumer in the example shown in Fig. 3 is obtained as follows:

- Consumer 1: $466.50 \times (312.08/518.16) = 280.97$
- Consumer 2: $466.50 \times (155.50/518.16) = 140.00$
- Consumer 3: $466.50 \times (50.58/518.16) = 45.54$

*D. Scheduling of Energy Consumption*

With the proposed pricing scheme (including the allocation of group price), the remaining challenge is how to schedule energy consumption of each consumer to maximize its saving.

Unlike the traditional grid, thanks to the bi-directional communication links shown in Fig. 1, the gateway can disseminate to all consumers in the smart grid the information on the consumption status of not only the whole group but also individual consumers in real time. Based on this information, each consumer can adjust (i.e., schedule) its energy consumption for a time

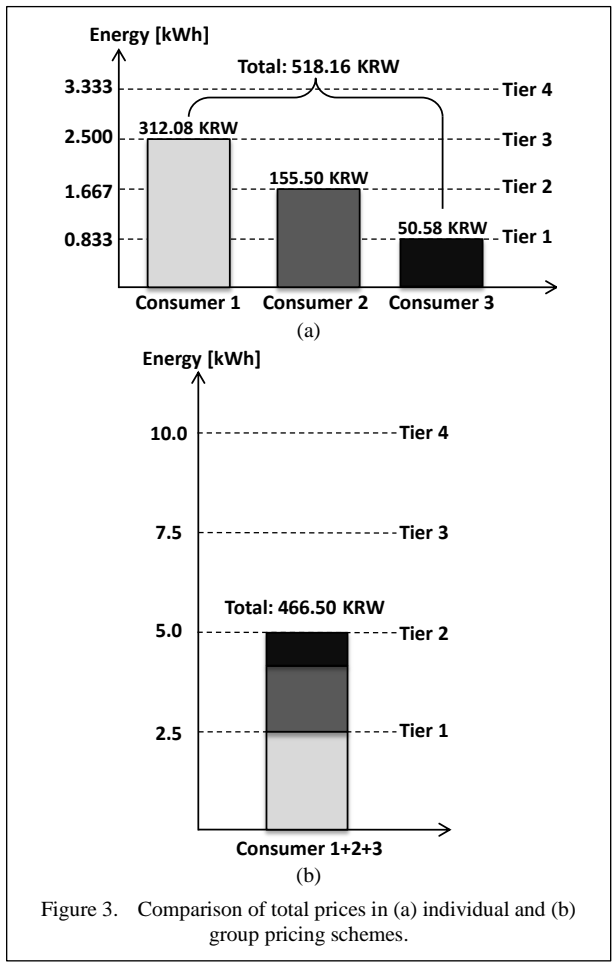

Figure 3. Comparison of total prices in (a) individual and (b) group pricing schemes.

period, typically longer than the period of time slot (e.g., a day).

Note that the implementation of energy consumption scheduling is beyond the scope of this paper and deserves a separate publication. Interested readers are referred to [4] and the references therein (especially [11]) for the game-theoretic approaches for this topic.

## III. CONCLUDING REMARKS

In this paper we have reported the current status of our research on designing pricing schemes based on progressive tariff and consumer grouping for a coexistence scenario in the migration to a future smart grid. To design a pricing scheme which is compatible with the existing progressive scheme in the traditional grid but can flatten the overall demand over time and thereby reduce peak load, we have proposed the application of a progressive tariff over a time slot to a group of consumers in a collective manner. We have also discussed the issue of price allocation under the proposed pricing scheme and provided an example of proportional allocation.

It is worth mentioning that with the proposed approaches, the static progressive tariff can be converted to a viable pricing scheme for a smart grid that is connected to a traditional electricity grid, which can provide monetary incentives for consumers to shift their consumption from congested time slots.

Note that the design of demand side management schemes including demand response and more advanced price allocation strategy for the proposed pricing scheme are two major topics for further study.

ACKNOWLEDGMENT

This work was supported by the Centre for Smart Grid and Information Convergence (CeSGIC) at Xi'an Jiaotong-Liverpool University (XJTLU).